\documentstyle[11pt,newpasp,twoside,epsf]{article}
\def\edcomment#1{\iffalse\marginpar{\raggedright\sl#1\/}\else\relax\fi}
\reversemarginpar

\begin{document}
\title{Chemical and Thermal History of the Intracluster Medium}
 \author{Hans B\"ohringer}
\affil{Max-Planck-Institut f\"ur extraterrestrische Physik,
Garching, Germany}

\begin{abstract}
Clusters of galaxies can be seen as giant astrophysical laboratories
enclosing matter in a large enough volume, so that the matter composition
can be taken as representing the composition of our Universe. 
X-ray observations allow a very precise investigation of the physical
properties of the intracluster plasma allowing us to probe
probe the cluster structure, determine its total mass, and
measure the baryon fraction in clusters and in the Universe as a
whole. We can determine the abundance of heavy elements from O to Ni
which originate from supernova explosions and draw from this
important conclusions on the history of star formation
in the cluster galaxy population. From the entropy structure of
the intracluster medium we obtain constraints on the energy release
during early star bursts. With the observational capabilities of
the X-ray observatories XMM-Newton and Chandra this field of research
is rapidly evolving. In particular, first detailed observations of the intracluster 
medium of the Virgo cluster around M87 have provided new insights. 
The present contribution gives an account of
the current implications of the intracluster medium observations,
but more importantly illustrates the prospects of this research for
the coming years.

\end{abstract}

\section{Introduction}

Clusters of galaxies, the largest well defined objects, 
are the largest well characterized astrophysical laboratories
at our disposal, except for the Universe as a whole.
The clusters masses range
from less than $10^{14}$ to several $10^{15}$
M$_{\odot}$. This mass range is extended down to about several $10^{12}$ 
M$_{\odot}$ by galaxy groups which can be considered
as small scale versions of clusters.
Hundreds to Thousands of cluster galaxies are making up
merely a few percent of the total cluster mass while more mass is in the gaseous
intracluster medium (ICM). The dominant fraction is Dark Matter.
Being formed through gravitational collapse 
where gravity acts on all forms of matter equally, 
galaxy clusters contain to a good approximation a composition of
matter well representative of the Universe as a whole (e.g. White et
al. 1993).

In this contribution I will focus on the information we can gain on
galaxy evolution from the study of the chemical composition and the
thermodynamic structure of the ICM. One of the astronomically most interesting epochs
in the history of our Universe was that of the most intense star and galaxy
formation at redshift between 2 and 4 
(e.g. Madau et al. 1998). The traces
of what happened then can be found in the ICM today:
i) we observe that the ICM is enriched by metals which must have been
synthezised by supernovae in the cluster galaxies mostly in these
early star formation epochs; ii) we further observe in the entropy
structure of groups and clusters of galaxies the effect of an early
heating of the intergalactic medium by energy release in the star 
formation epochs.

The ICM is a hot plasma with temperatures of
several ten Million degrees which has its thermal radiation maximum in
the soft X-ray regime. It is a fortunate coincidence that it is the
same wavelength region in which we can use the current technology of
imaging X-ray telescopes. Therefore we know most about the ICM 
through X-ray astronomy. We have furthermore two new satellite
X-ray observatories, ESA's XMM-Newton mission and NASA's Chandra
mission, with greatly improved spectral and imaging capabilities, which
are providing a overwhelming new insight into the physics of the
ICM. 
In this contribution I will use a varying scaling of physical parameters
the Hubble constant (depending on the sources) quoted which will be labled
by the scaling parameter $h$,
e.g. $h_{70} = H_0 / 70$ km s$^{-1}$ Mpc$^{-1}$.

\begin{figure}[ht]
\plotone{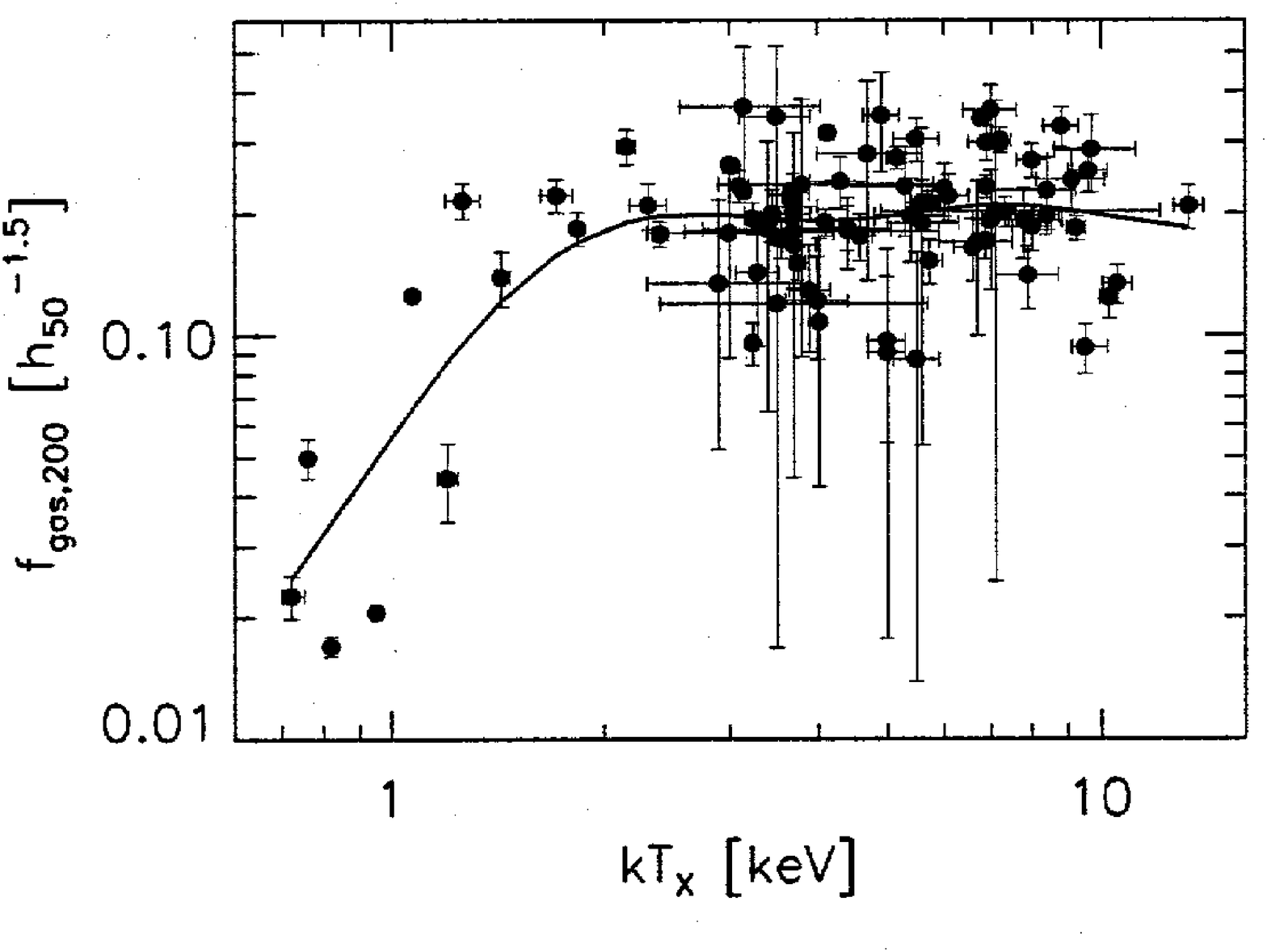}
\caption{Gas mass fraction in units of $h_{50}^{-1.5}$ for
106 of the brightest clusters of galaxies detected by ROSAT in X-rays
(from Reiprich 2001).
}
\end{figure}

\section{The mass fraction of the intergalactic gas}

\begin{figure}[h]
\plottwo{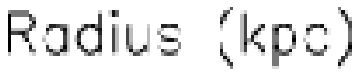}{Boehringerfig2b.ps}
\caption{{\bf Right panel:} Temperature profile of the cluster
A1413 determined from
XMM-Newton spectroscopic observations by Pratt \& Arnaud (2002). This
temperature profile stretches out to the largest radii in a
cluster analysis reached so far, to 0.6 of the virial radius.
{\bf Left panel:} Gas mass fraction in A1314 from Pratt \& Arnaud (2002).}
\end{figure}

A prime characteristic of the matter budget is the fraction
of baryonic matter with respect to the Dark Matter.  
The gas mass as well as the total mass can be determined from the
gas density and temperature distribution of the ICM 
(assuming hydrostaic equilibrium) which is in turn
determined from X-ray imaging and spectroscopy.
Fig. 1 shows the gas mass fractions
determined for a sample clusters observed with
ROSAT and ASCA (Reiprich 2001, Reiprich \& B\"ohringer 2002). The data
scatter around a value of 18-20\% (for $h = 0.5$; or 12\% for $h = 0.7$). 
The large error bars in the mass are due to
uncertainties in the temperature and temperature profiles. 
The gas mass fraction seems to be quite constant for massive clusters but
decreases towards the galaxy groups.
With the capability of spatially resolved spectroscopy with XMM Newton
and Chandra the precision of such measurements greatly improves. 
One of the best recent examples is the XMM study
of the galaxy cluster A1413 shown in Fig. 2 (Pratt \& Arnaud 2002). Here
the gas mass fraction has been determined out to a radius close to
the assumed virial radius of the cluster. The resulting gas mass
fraction is $0.2 (\pm 0.02) h_{50}^{-1.5}$. Similar
results with a more limited radial range have been obtained
with the CHANDRA observatory (e.g. Allen et al. 2001, Ettori et
al. 2002).

Assuming that most of the unseen matter is 
non-baryonic, we can determine the baryon fraction
of the total cluster mass by summing the observed gas and galaxy
mass (the latter making up about 2-3\% of the
total mass) finding values around $12\% h_{70}^{-1.5}$. Taking this 
fraction as representative of the Universe together
with the matter density of the Universe as provided by the cosmological
''concordance'' model (e.g. Turner 2002) we get $\Omega_b h^2 = 0.191$.
This is consistent with the value from nucleosynthesis for the observed
primordial deuterium to hydrogen ratio (e.g. O'Meara et al. 2001) 
giving $\Omega_b h^2 = 0.020$  and the analysis of microwave background 
observations providing $\Omega_b h^2 = 0.0224$ 
(for $h = 0.7$ and $\Omega_m = 0.3$).
This general consistency is also a confirmation of the mass and gas
mass fraction determination in clusters.

\section{Observations of the metallicity of the intracluster medium
before XMM-Newton}

\begin{figure}[h]
\plottwo{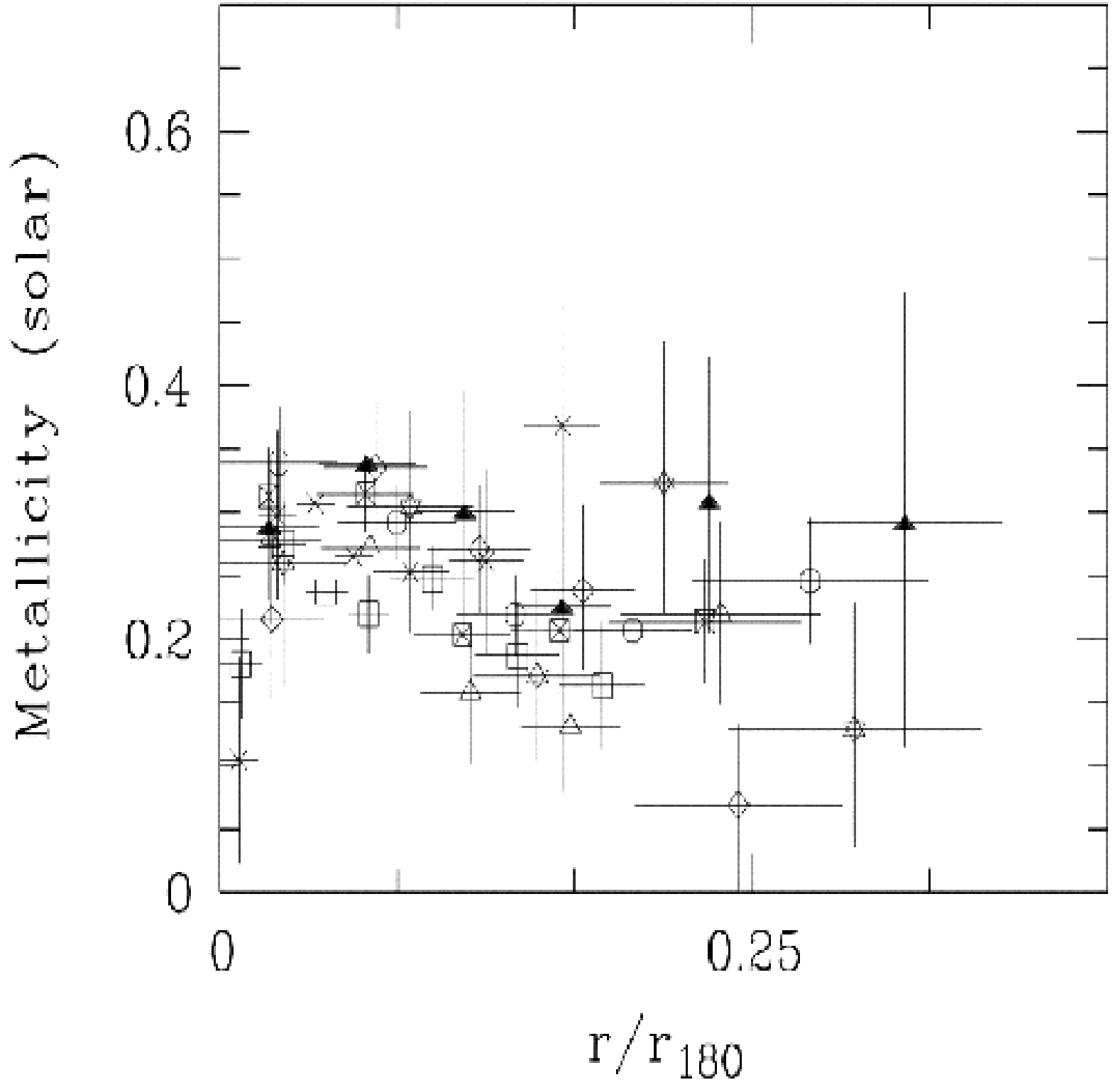}{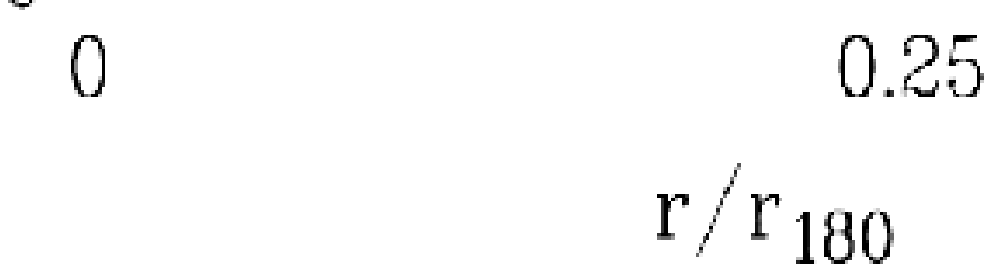}
\caption{
Iron abundance profiles of non-cooling flow clusters (left) and cooling flow
clusters (right) as a function of the scaled radius expressed in units of 
the radius at
a mean mass overdensity of 180 over the critical density of the Universe
(De Grandi \& Molendi 2001).}
\end{figure}

The first studies of the metal abundance distribution
with ASCA and BeppoSAX observations showed a central 
increase of the Fe abundance in clusters
with so-called cooling flows (e.g. Matsumoto et al. 1996,
Fukazawa et al. 1998, Finoguenov et al. 2000, 2001), while 
cluster without a dense core and without a central cD galaxy 
display flat iron profiles as shown in Fig. 3 (De Grandi \& Molendi 2001).
These results lead to the tentative implication that the central Fe
abundance increase is connected to recent iron production in the
central early type galaxies. The principle sources of the heavy elements
are the two types of supernovae, type II (core collapse
supernovae) and type Ia (thermonuclear exploding white dwarfs).
Only the supernovae of the type Ia have been observed in the early
type galaxies. Therefore the increased abundance of iron in the
cluster centers must come from the enrichment by SN type Ia. This can be tested
by looking at the relative
abundances of different elements produced in different ratios by SN Ia
ans SN II. Using ASCA results
Fukazawa et al. (1998) and Finoguenov et al. (2000, 2001a)
showed a clear trend for an increase in the Si to Fe ratio with radius,
implying that SN II with relative high Si yields are dominating in the 
non-central cluster regions.

\section{Metal abundance determination in the X-ray halo of M87 with
  XMM-Newton}

The giant elliptical galaxy, M87, in the center of the northern part
of the Virgo cluster is the nearest X-ray luminous galaxy cluster
center, providing the
highest signal-to-noise for X-ray spectral analysis and the
best angular resolution for the study of the dense central ICM.
In addition the temperature of the X-ray halo of M87 is
with values in the range from 1 to 3 keV in a regime
where the spectra are very rich in emission lines. For its
importance M87 was one of the verification targets of the
XMM-Newton mission. 

To determin element abundances
the temperature structure of the ICM has to be known
with high precision. In particular, the M87 halo was previously
believed to harbour a classical cooling flow,
proposed to display a complex multi-temperature structure with a wide
range of temperatures (e.g. Nulsen 1986, Thomas et al. 1987, Fabian
1994) which can easily lead to ambiguous abudnace results
(e.g. Buote 1999). Therefore
the first analysis was a careful determination
of the temperature structure of the M87 halo (Matsushita
et al. 2002). It turns out that the classical cooling flow picture
does not apply (Fig. 4, B\"ohringer et al. 2001, 2002, Moldendi
et al. 2001, Molendi 2002) and that the gas is very close have a single
temperature phase locally, but shows a monotonic decrease in
temperature towards the center (Matsushita et al. 2002) which greatly
facilitates the data interpretation.

\begin{figure}[h]
\plotone{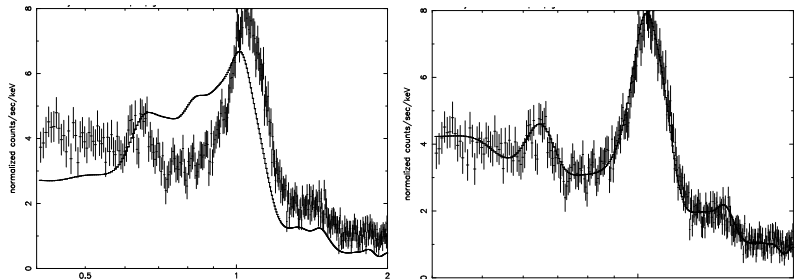}
\caption{The observed Fe L line feature in the X-ray halo of M87
(around a radius of 2 arcmin) compared to a single temperature (right) and a
classical cooling flow model (left; B\"ohringer et al. 2002).
}
\end{figure}

\begin{figure}[h]
\plottwo{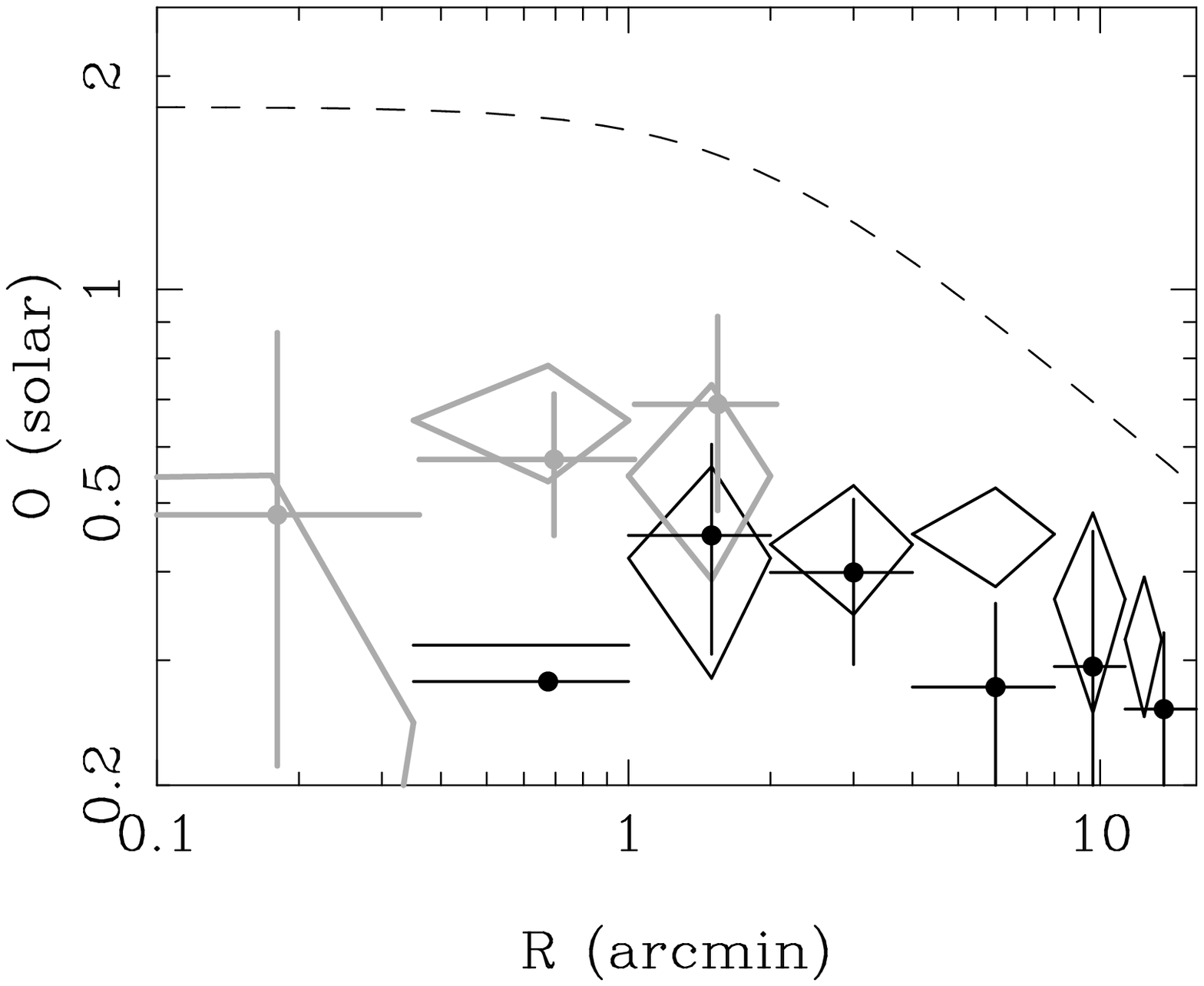}{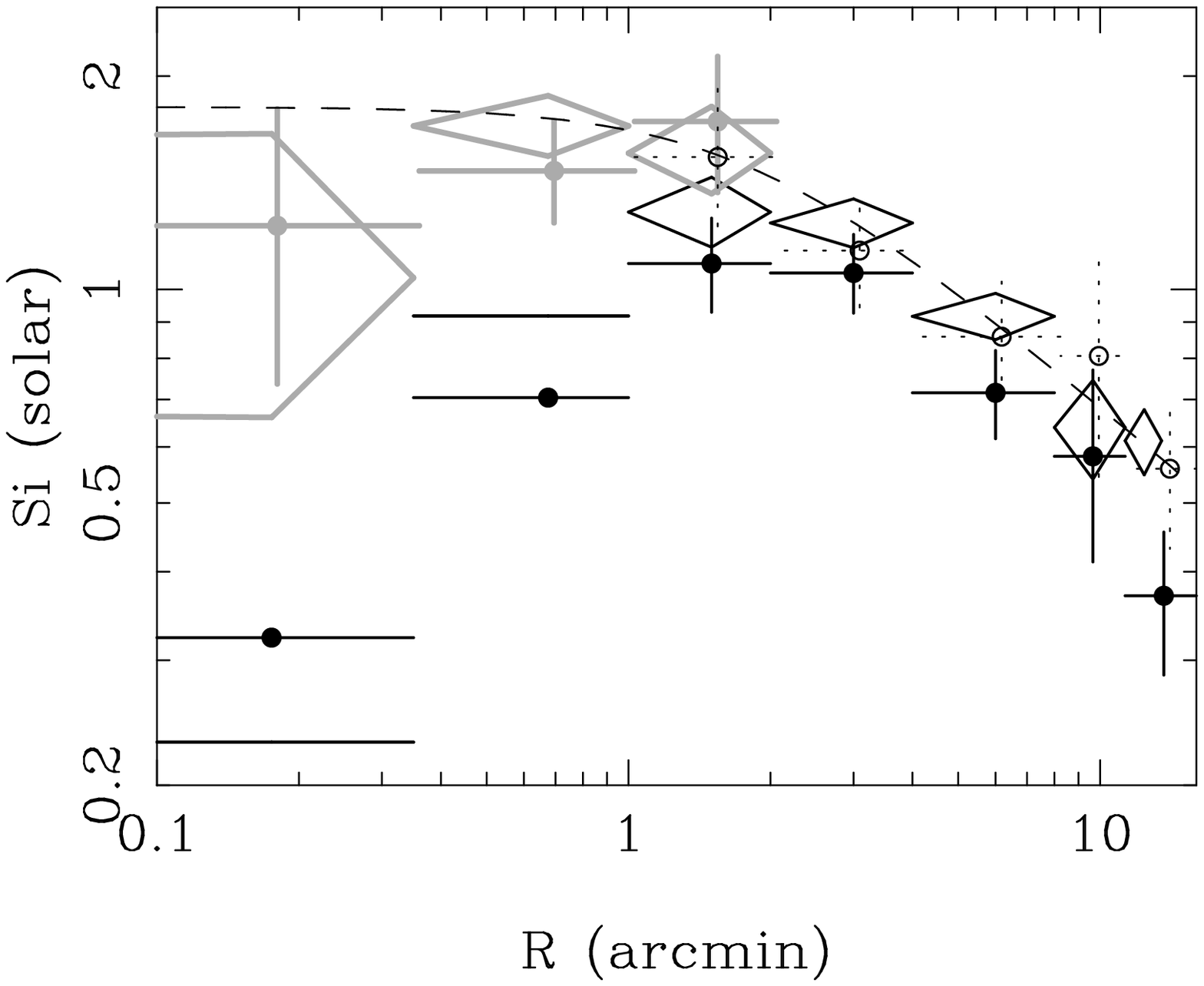}
\plottwo{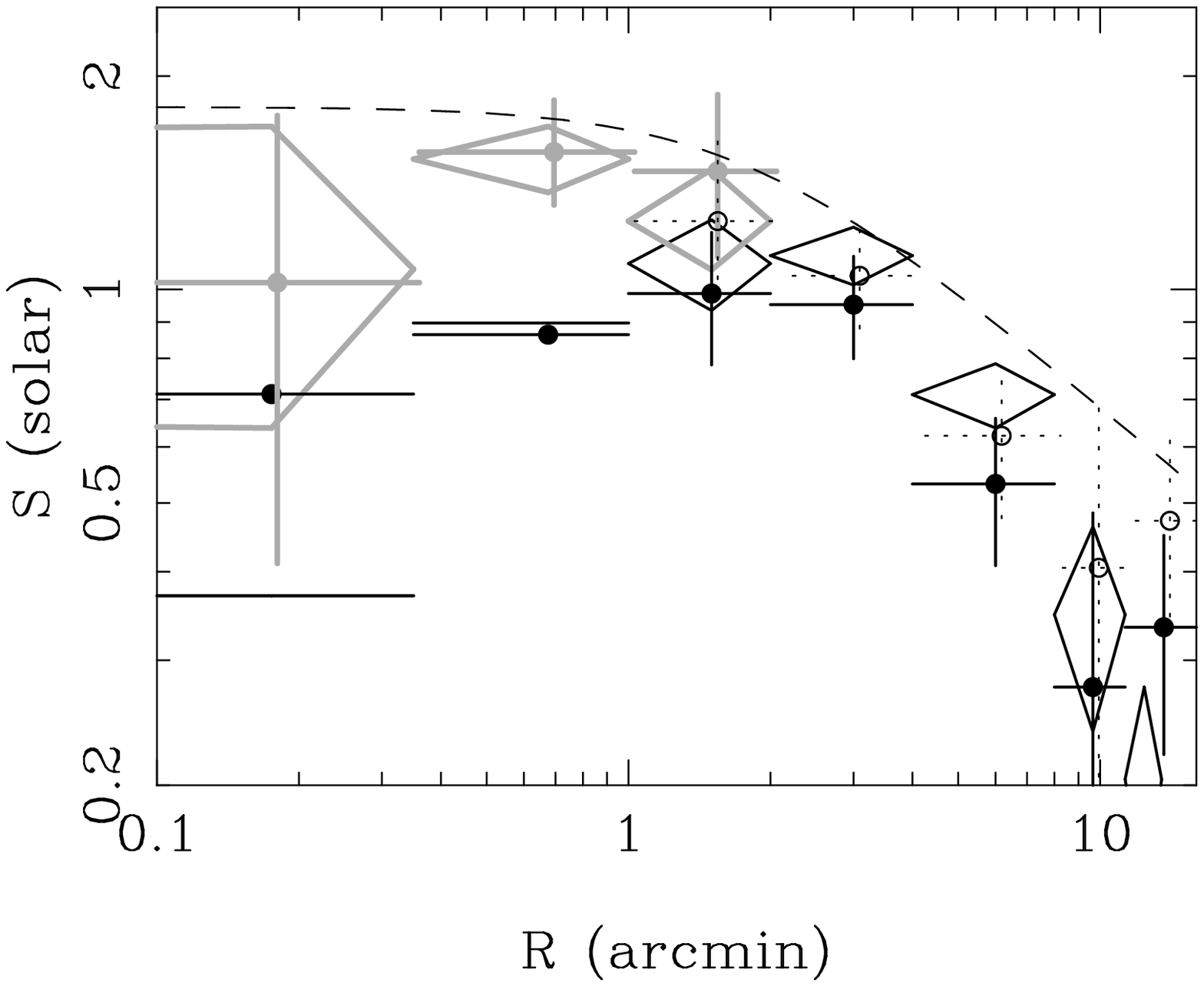}{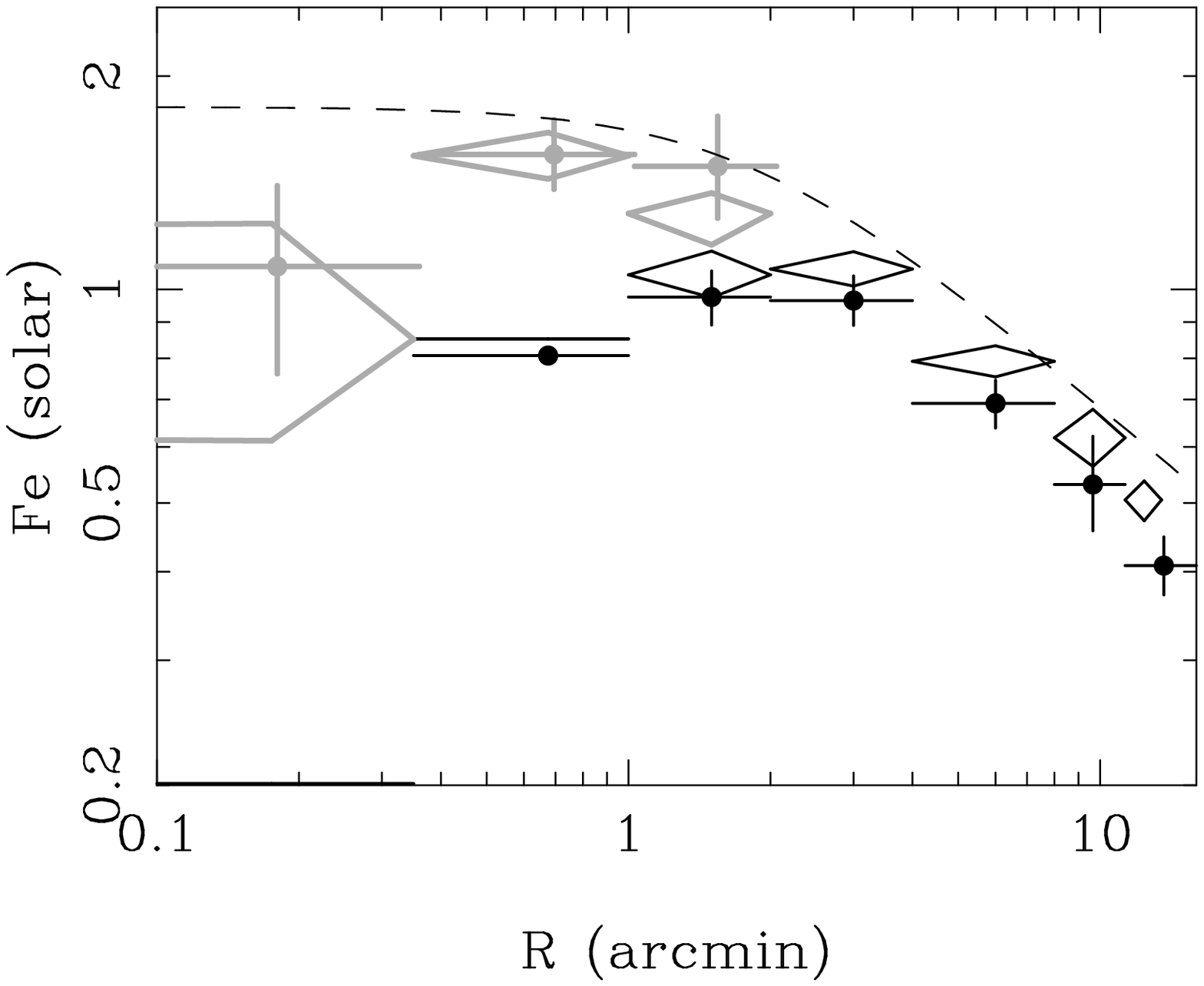}
\caption{
O, Si, S and Fe abundance profiles in M87 (Matsushita et al. 2003a).
The grey symbols came from a two-temperature model (grey symbols)
for the central region where the disturbance of the central AGN
may not be perfectly subtracted. These results are more reliable
than the simple one-temperature treatment (black symbols). The data
are obtained from the two XMM detectors EPN (crosses) and EMOS (diamonds).}
\end{figure}

Abundance profiles in M87 determined (for deprojected spectra) 
for the elements O, Si, S, and Fe are
shown in Fig. 5 (Matsushita et al. 2003a).
We clearly see a different behavior of the different
elements. While O shows a relatively flat profile, the profiles of Si
and Fe decrease significantly with radius (in approximately the same
way) and S seems to fall off even more steeply. This is partly what we
could have expected: oxygen which is produced essentially only by SN
type II at early epochs and which should be well mixed shows the more
even distribution. The strong increase in Fe and Si is then due
to the enrichment by SN Ia.
It comes somewhat as a surprise, however, that the profiles of Fe and
Si are practically running in parallel. 
We should have expected a stronger increase of Fe compared
to Si, because in the classical models the SN type Ia which are
responsible for the central increase of the abundance profiles, should
produce predominantly iron and less Si.

\begin{figure}[h]
\plottwo{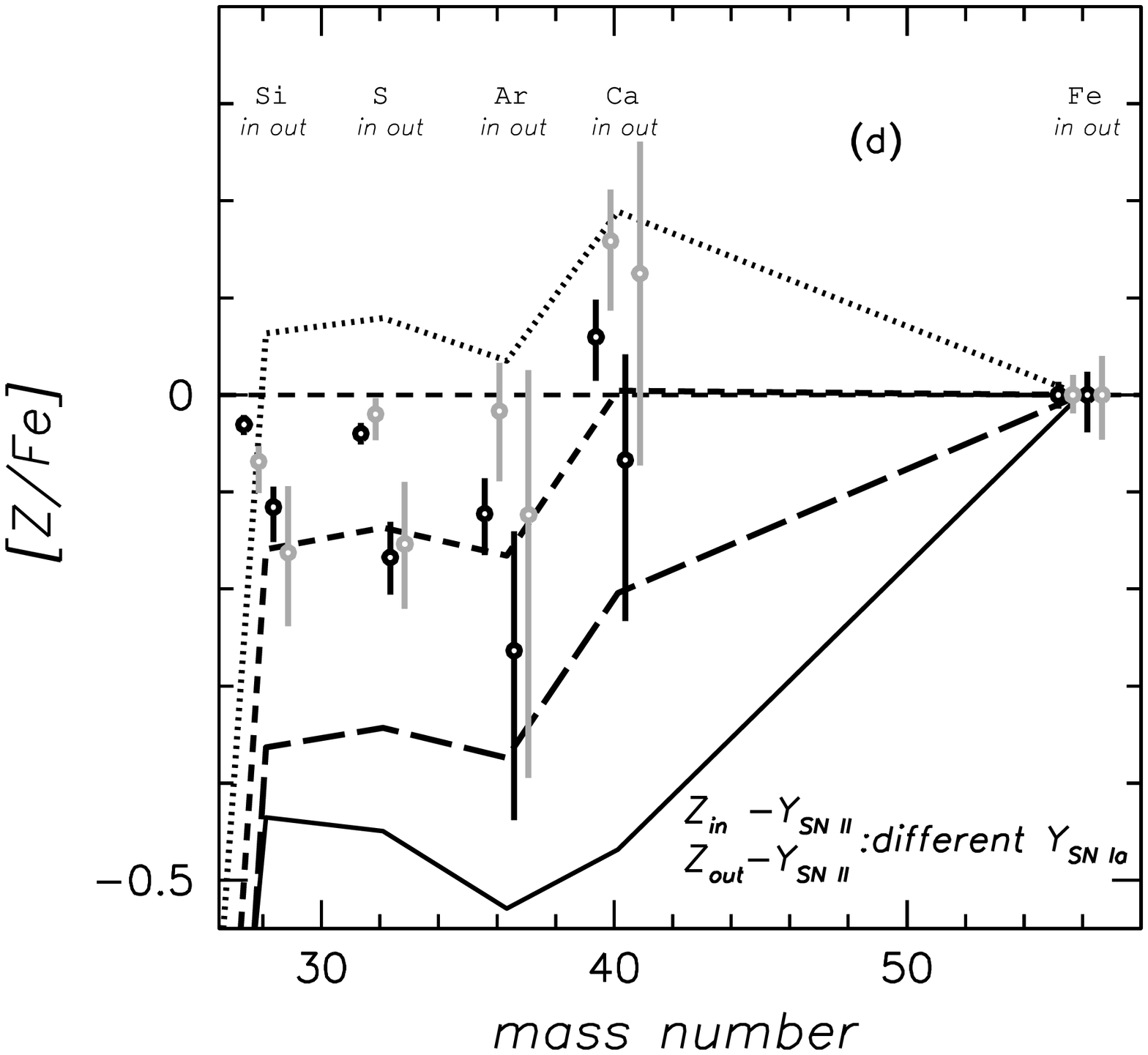}{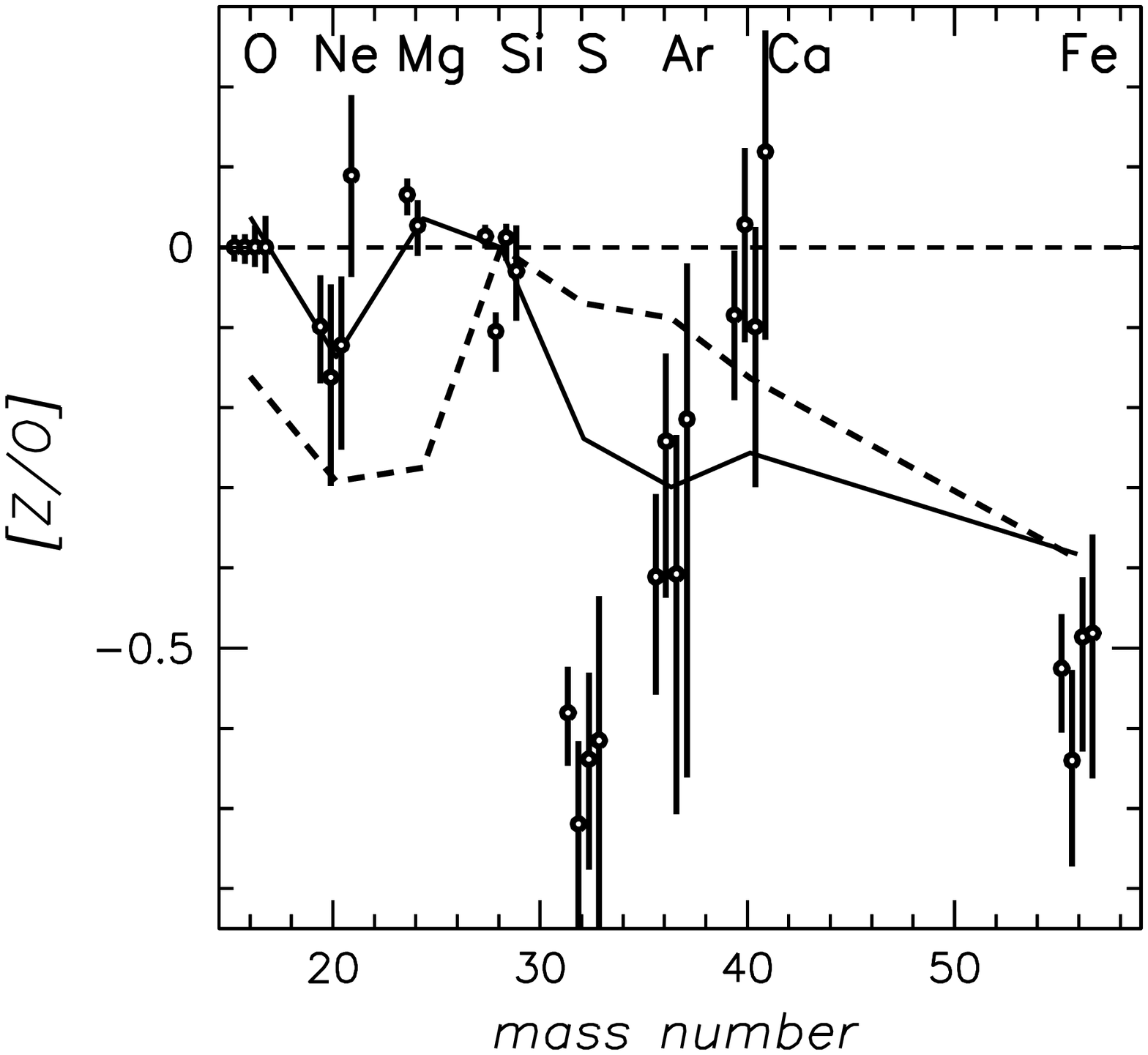}
\caption{
The contribution of SN type Ia (left) and SN II (right)
to the abundance pattern in M87 (Finoguenov et al. 2002).
Also shown is a comparison with nuclear synthesis model calculations:
{\bf left panel:} the lowest model curve is the classical SN Ia model
of a fast explosion by Nomoto et al. 1984, the other three curves are
slow deflagration-detonation models with the higher curves having
decreasing deflagration speeds and less complete burning (Nomoto et al. 1997).
{\bf Right panel:} Shows a comparison with SN II models by Nomoto
et al. 1997 (upper curve) and Woosley and Weaver (lower curve).}
\end{figure}

To investigate this further the behavior of all the elements was
studied by Finoguenov et al. (2001b) in two radial bins (1 to 3 arcmin
and 8 to 16 arcmin) in M87. The result was 
deconvolved into the contributions of the two different types of supernovae
under the assumption that the central abundance increase is 
entirely due to SN type Ia. One can then subtract the
outer abundance pattern from the inner one to obtain the SN Ia yields. 
Alternatively on can use oxygen as a tracer of the contribution of 
SN type II and subtract
from the inner abundances a weighted outer abundance pattern
such that the O abundance is zero. (see
Finoguenov et al. 2002 for details). The results of this
procedure are shown in Fig. 6.

These deconvolved abundance patterns are compared to theoretical
models in the Figure. For the SN type Ia yields models by Nomoto et
al. (1997, see also Iwamoto et al. 1999) are used for the comparison.
The model with the lowest yields for the $\alpha$-elements compared to
the Fe group elements is the classical Nomoto et al. (1984) W7 model
of a fast SN Ia explosion which has been widely used for chemical
evolution models. The relative yields inferred from the present
observations show much larger $\alpha$-element yields more in line
with the slow deflagration-detonation models featuring a less complete
burning of the $\alpha$-elements. One also notes a second order result
that the two patters of SN type Ia for the inner and outer region do
not perfectly agree, but the outer region seems to feature
relatively lower yields for the $\alpha$-elements. 
This may be an indication of a secular variation of the nature of
SN Ia supernovae, with a smaller yield (less complete burning and less
bright light curves) for Fe group elements for a younger stellar population
and a relatively higher yield of $\alpha$-elements in the very old stellar 
populations of elliptical galaxies, like M87. This is in line with
the statistics that faint SN Ia are preferentially detected in early
type galaxies (Ivanov et al. 2000). This was investigated in more detail by Matsushita
et al. (2003a, 2003b) in M87 and Centaurus confirming this picture. 
An analysis of the enrichment times involved in the production
of the central metal abundance enhancement is described in B\"ohringer
et al. (2003b) where it is found that the enrichment time in the very inner zone
of M87 (inside the scale radius of about 10 kpc) is about 2 - 3 Gyrs while
in the enrichment time integrated out to 50 kpc has to be of the order of 
10 Gyrs. This enrichment age difference that roughly applies to the 
two radial zones shown in Fig. 6 can quite well be related
to a significant difference in the stellar population and enrichment
histories of the inner and outer region of the M87 halo.

\section {Entropy structure of galaxy groups and clusters}

\begin{figure}[h]
\plottwo{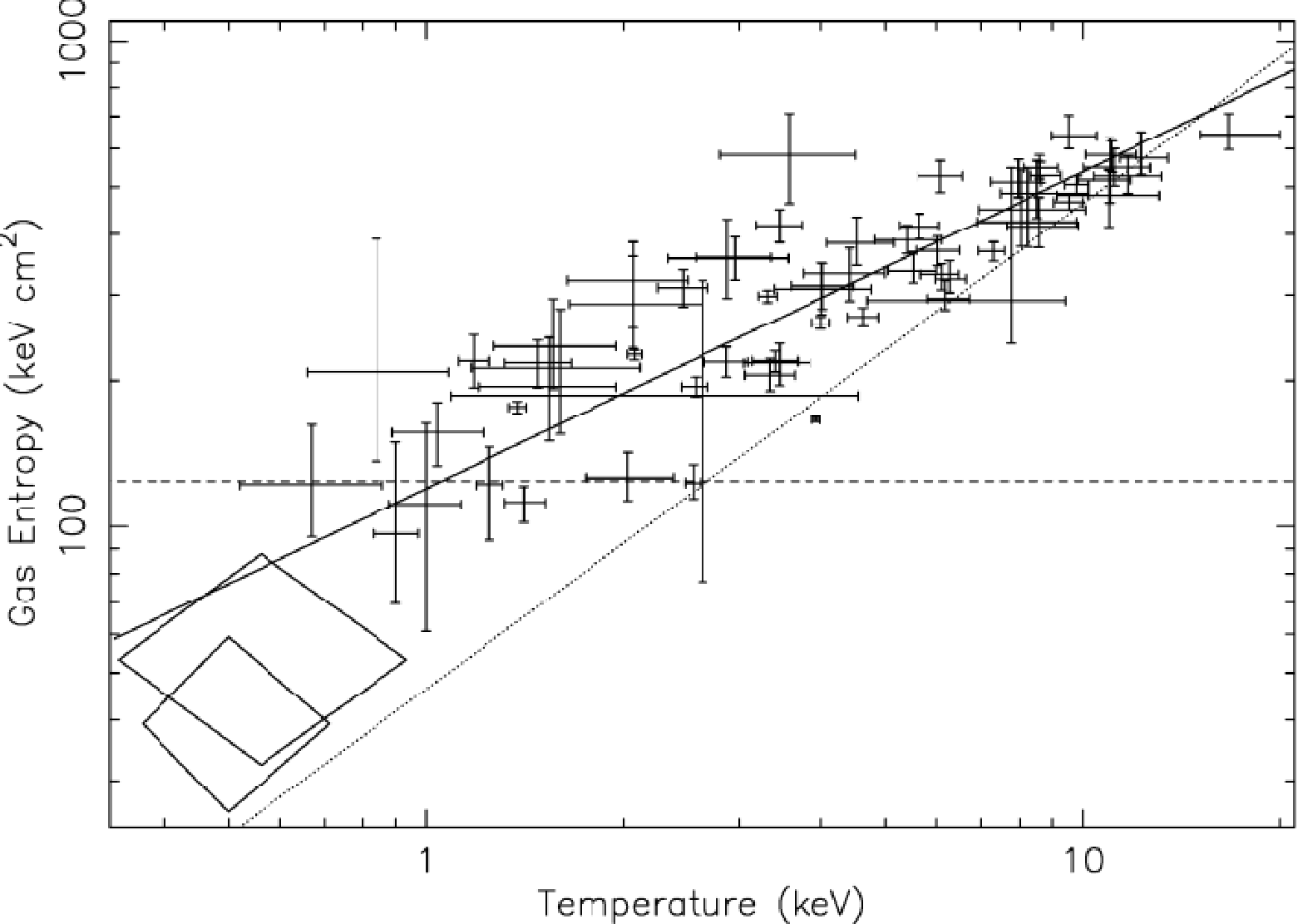}{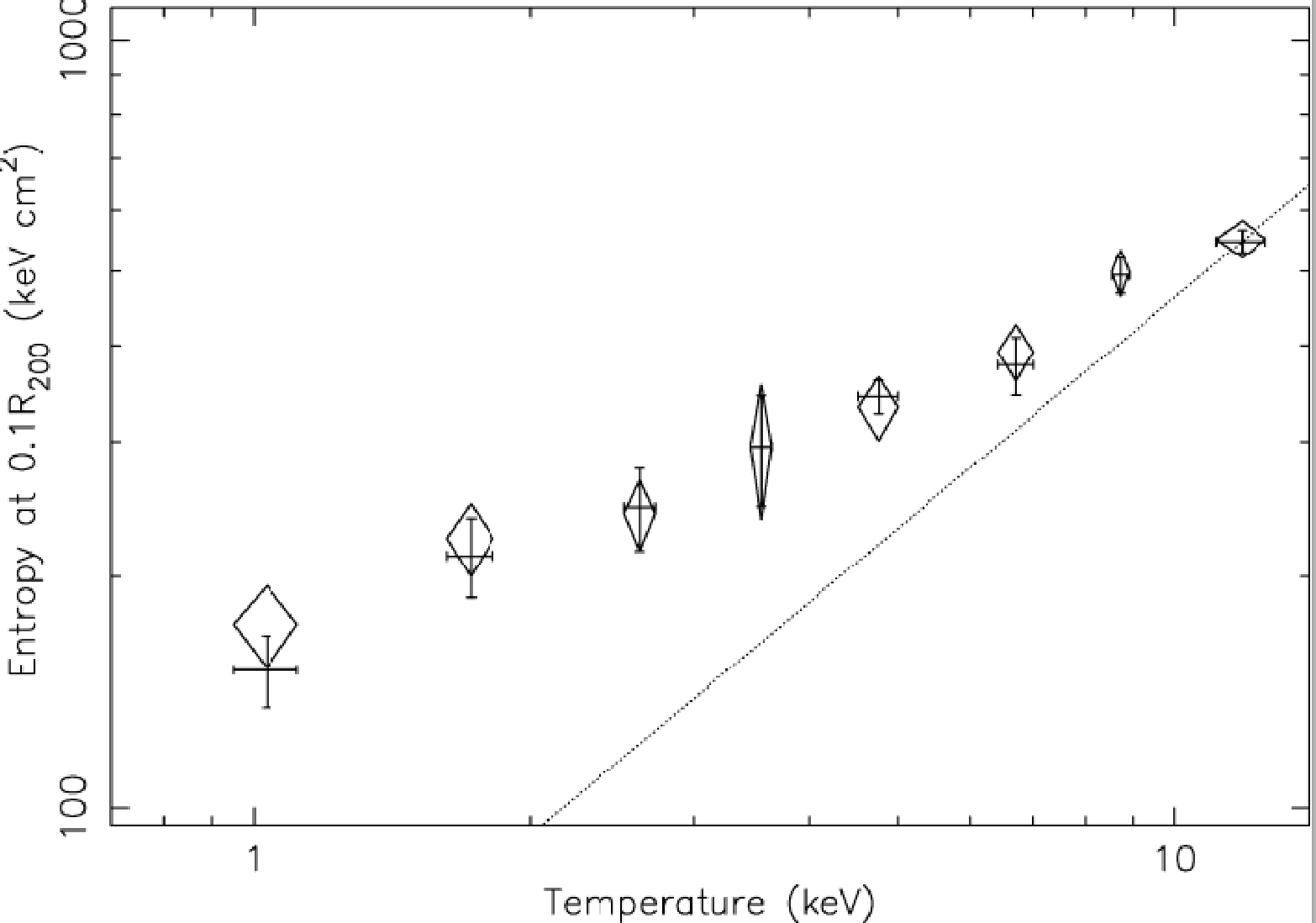}
\caption{
Entropy at a radius of 0.1 $r_{200}$ as a function of ICM temperature
for groups and clusters of galaxies (left) including two galaxies (diamonds).
Same data grouped into bins of 8 objects excluding the galaxies (right).
The solid line shows the best fit and the dotted line the self-similar
slope of 1 (Ponman et al. 2003)}
\end{figure}

The formation of dark matter halos from primordial overdensities is a
self similar process. The structure of the groups and clusters of 
galaxies is essentially determined by the potential shape of the dark matter
halos in which they are embedded. Therefore many of the
properties of clusters are well characterized by self-similar scaling
relations, like the mass - ICM temperature relation which is predicted to 
have a logarithmic slope of 1.5 which is approximately observed 
for massive clusters (e.g. Finoguenov
et al. 2001b). A closer comparison of groups and
clusters shows, however, that the predicted scaling relations are not
always very well observed - the temperature - X-ray luminosity
(bolometric luminosity) relation is a good example where the observed slope
is steeper than predicted (e.g. Navarro et al. 1996).

The possible reason for this is revealed in the diagnostics pioneered
by Ponman et al. (1999). Shown in Fig. 7 the entropy at a radius
of 0.1 $r_{200}$ (an inner radius avoiding cooling flow regions
scaled to the virial radius $r_{200}$ of the cluster)
as a function of system temperature.
The hot gas temperature and the entropy is in general the result of
shock heating of the gas when the cluster is subsequently assembled by
a series of substructure mergers. The shock strength should thereby be
proportional to the depth of the gravitational potential in which the
accreted matter settles which is itself proportional to the
temperature. Thus we expect that the gas entropy (expressed here with
the specific definition of $S = T/n^{2/3}$) is proportional to the
shock strength and thus to the temperature. The shallower slope
observed is indicative of an extra source of heat
which affects specifically the low
temperature systems. Such an effect is easily explained if the heat
input is proportional to the gas mass, that is a mechanism
that provides a source of fixed specific thermal energy. An heat
input proportional to the stellar population mass that is heating from
star formation would fullfil this requirement.
The extra entropy in the central regions amounts to typical values of 100 to
200 keV cm$^2$. 

One obvious source of a fixed amount of specific energy is the energy
released by those supernovae which also contribute to the enrichment of
the intracluster medium. From the above considerations on the metal
enrichment of the intracluster medium we can also calculate the energy
that is released by the same supernovae. Assuming that not much energy
is dissipated locally in the galaxies where the supernovae explode but
that most of it is transported out by a galactic wind, one finds that
about 0.5 to 0.75 keV per proton is available (see e.g. also
Finoguenov et al. 2001a) for an early heating
of the intergalactic medium which could result in the observed entropy
excess. Llyod-Davies \& Ponman (2000) have attempted to determine this
extra energy through the binding energy, finding values of about 0.5
keV per proton.
Knowing the energy and entropy excess we can actually calculate at
which environmental density this entropy excess was created with a result
of about $ 1 - 2~ 10^{-4}$ cm$^{-3}$. This gas density is reached in the
cosmic mean gas density at redshifts of about 6 to 7 and for an object
that is at turn-around and starting to collapse at a redshift of 3 to 4.
The latter is a plausible number for the peak of star formation 
in a protocluster environment. In addition cooling processes are operative 
in the group and cluster centers which have to be taken into account.

This rough picture is tested in simulations by a number of authors.
Some have shown that the amount of
energy needed at the typical time of most of the star formation to
produce the entropy excess is much higher than what can be expected
from the energy release of supernovae, even if all this energy is
dumped into the intracluster medium with negligible spontaneous energy
dissipation (e.g. Wu et al. 2000, Borgani et al. 2002).
Voit \& Bryan (2000) as well as others
(Bryan 2000, Pearce et al. 2000, Muanwong et al. 2001, 2002, Wu \& Xue 2002,
Voit et al. 2002 Voit \& Ponman 2003)
have shown that the cooling and condensation of the coldest
gas in groups and clusters can produce the observational effect. But
it would imply a much to high degree of conversion of gas into stars
(e.g. Balogh et al. 2001).
Thus recent models have been more successful by combing the effect of
cooling and feedback heating, e.g. by Finoguenov et al. (2003) applying 
a homogeneous specific energy input of 0.75 kev per
particle at a redshift of $z \sim 3$ with additional cooling.

\section {Summary and conclusions}

The above results of XMM-Newton and CHANDRA observations 
demonstrate that we are starting to gain precise knowledge
about the structure of galaxy clusters and the physical state of
the ICM. In particular study of the the abundance pattern and
the entropy structure in the ICM of clusters and groups   
is shedding light onto the supernova yields, the metal enrichment history
in relation to the structural evolution of the cluster, and the star
formation evolution. The detailed solution of most of the 
questions rised here are still to come. 
Therefore we will be witnessing a great time for X-ray astronomy providing
substantial contributions to the major questions of astrophysics and cosmology.

\acknowledgments{
I like to thank my collaborators Kyoko Matsushita, Alexis Finoguenov,
Ulrich Briel, Stefano Borgani for the fruitful joint work and the many
discussions. I also like to thank the XMM-Newton team at MPE and
outside for support and for making these beautiful results possible.
XMM-Newton, an ESA mission, is funded by ESA Member States and NASA. 
The XMM-Newton project is supported by the Bundesministerium f\"ur
Bildung und Forschung, Deutsches Zentrum f\"ur Luft und Raumfahrt, The Max-Planck Society and
the Haidenhain-Stiftung.
}

\end{document}